\documentclass{cimento}

\usepackage{graphicx,amsmath,amssymb,amsfonts,latexsym,color,dcolumn,bm}
\usepackage{braket}
\usepackage{makeidx}
\usepackage{multirow}
\usepackage[dvipsnames,svgnames,table]{xcolor}
\usepackage{graphicx}
\usepackage{epstopdf}
\usepackage{ulem}

\newcommand{\be}{\begin{equation}}
\newcommand{\ee}{\end{equation}}
\newcommand{\bea}{\begin{eqnarray}}
\newcommand{\eea}{\end{eqnarray}}
\newcommand{\ad}{a^{\dag}}

\title{Complementarity and light modes}
\author{R.~Menzel  \from{ins:x}}
\instlist{\inst{ins:x} University of Potsdam, Institute for Physics and Astronomy\\
Karl-Liebknecht-Str. 24/25, 14476 Potsdam, Germany}

\PACSes{\PACSit{--.--}{\dots}
\PACSit{--.--}{\ldots}}

\begin{document}

\maketitle

\begin{abstract}
In quantum physics we are confronted with new entities which consist indivisible of an energy packet and a coupled wave. The complementarity principle for certain properties of these quantum objects may be their main mystery. Photons are especially useful to investigate these complementary properties. A series of new experiments using spontaneous parametric down conversion (SPDC) as a tool allowed the detailed analysis of the physical background of this complementarity and offers a new conceptual perspective. Based on these results a straightforward explanation of these sometimes counter-intuitive effects is given. 
\end{abstract}

\section{Introduction}
Complementarity is one of the most fundamental and important principles of quantum physics. It was mentioned the first time by Niels Bohr in 1927 (at the Fermi conference in Como) as a consequence of the uncertainty of certain pairs of quantum physical parameters and published one year later \cite{ref:Bohr}. The most prominent pair of such parameters are position and momentum. Other pairs, as e.g. electric and magnetic field strengths of photons, can be derived from them. But complementarity is also a consequence of the wave-particle-duality and as such very widely investigated in quantum optics. The most prominent example of this duality appears in the double slit experiment with single photons, where we obtain interference structures in the light pattern behind the double slit if no which-slit information is available and no interference (but diffraction) if the knowledge about which slit the photon has passed is available \cite{ref:Wooters}.

In his famous lectures on physics Richard Feynman wrote in 1964 that the wave-particle dual behavior contains the basic mystery of quantum mechanics \cite{ref:Feynman}. Later he even pronounced it more by saying it may be even the only mystery of quantum mechanics. In 1984 Marlan Scully, Berthold Englert and Herbert Walther stated that: Complementarity is deeper and more general and fundamental to quantum mechanics than uncertainty \cite{ref:Scully1}.

Although the present theoretical concepts of quantum physics provide a complete set of recipes to calculate all results which have so far been observed with quantum systems these experimental results are sometimes quite counter-intuitive. Therefore it seems to be worthwhile to investigate the physical background of complementarity in further detail to reach a better conceptual understanding. As it will turn out at the end of this manuscript it seems possible to identify a more or less hidden physical entity as responsible for these astonishing effects. Its action is completely included in our theoretical concepts in a very optimal way. But it is only indirectly observable: the quantum vacuum \cite{ref:MilonniVac}.

For the investigation of complementarity and for the detailed analysis of the wave-particle-duality photons as quantum objects are especially useful. In our everyday life at room temperature only about 1\% of the light modes are occupied by photons, 99\% are empty. So far the energy packets of photons can easily be separated with their related modes in contrast to the spectral range of radio waves. On the other side in the high-energy region the wave properties are very difficult to observe because of technical difficulties in this very short wavelength range.

As a result in the measuring process the energy packet of the photon is measured as a click. But the photon modes show interference and diffraction. Clicks are not observed at spots of distractive interference. On the other side even single photons can be detected in spatial regions which would never be illuminated without the diffraction, e.g. in the dark region behind the knife edge. Therefore it can be concluded that the single photon (as any other quantum object) consists of an energy packet and an associated wave which are indivisible. This duality has far-reaching consequences in the measuring process.

In general light or single photons as the constitutents of light are distributed \begin{bf} spectrally, temporarily, spatially and in their polarization\end{bf}. But usually the detection system of light is only sensitive in certain ranges of these distributions. As a result in the measurement process the single photons are detected only within the chosen distribution of frequency, time, space and polarization. These distributions define the modes of the detection system which are known from classical optics.\\

\begin{bf}
Only photons within the detection mode can be registered as clicks.
\end{bf}\\

As a result in each physical situation only a certain selection of the photons and thus only a certain selection of the reality of that quantum system is registered. As it will be shown below this selection determines the properties of the observed reality. In other words:\\

\begin{bf}
It is the measuring process which selects the properties of the quantum object in a certain physical situation.
\end {bf}\\

Of course photons cannot be detected if they are never generated. Thus the result of the measurement is the convolution of the modes of generation with the modes of detection. (In many cases the distributions of the generation process are much wider than the distributions of the detection process and therefore in these cases the detection modes play the dominant rule).

Regarding complementarity two fundamental aspects can be extracted from the non-separable energy-packet and wave nature of single photons (as for all of other quantum objects):\\

1)	The wave nature of the photon results in a number of uncertainty relations as e.g. for space and momentum. The spatial modes of single photons as used in quantum field theory are solutions of the classical Maxwell or Helmholtz equations. These modes exist in free space either occupied or not occupied by photon energy packets. The classical light mode with the smallest product of beam waist diameter and divergence is the TEM$_{00}$-Gauss-mode. The spatial uncertainty within the beam waist and the momentum uncertainty within the divergence fulfill exactly the Heisenberg uncertainty relation between space and momentum \cite{ref:Photonics}, which is no surprise because the underlying quantum theory is a wave description.\\

{\centering
\fbox{\parbox{\linewidth}{
\centering
\begin{bf}
The uncertainty relations describe the complementarity between the\\ involved parameters of the considered quantum object\\ (as e.g. the photon).\\
\end{bf}
}}}\\
\\

2)	The photon modes in consideration may show certain coherence properties. If we assume for simplicity the mode function system of Gauss-Laguerre and Gauss-Hermite modes the longitudinal and the transversal coherence lengths of these modes can be defined \cite{ref:Born-Wolf}. Therefore the question arises how coherent are single photons in these modes and how distinguishable are they. The coherence is usually characterized by the visibility V of the fringes of an interference experiment which is calculated from ${V = (C_{max}-C_{min})/(C_{max}+C_{min}) }$. ${C_{max}}$ and ${C_{min}}$ are the maximum and the minimum photon count rates of the fringes. The distinguishability can be measured in coincidence with the reference photon as a marker for one of the two paths the photon can take. The distinguishability D follows from ${D = (R_{path 1}-R_{path 2})/(R_{path 1}+R_{path 2}) }$. In this formula ${R_{path 1}}$ and ${R_{path 2}}$ are the coincidence count rates measured for the observed photon in path 1 and in path 2 in coincidence with a reference photon marking path 1. As example think of the double slit experiment there path 1 relates to slit 1 and path 2 relates to slit 2. From a fundamental quantum optical calculation follows that $D^2+ V^2$ equals in maximum 1 \cite{ref:Englert} and \cite{ref:Englert2}. This relation is also true for physical situations of uncertainties larger than Heisenberg's uncertainty relation would demand, as will be shown below.\\

{\centering
\fbox{\parbox{\linewidth}{
\centering
\begin{bf}
The relation between visibility and distinguishability is the second fundamental aspect of the complementarity principle for quantum objects.\\
\end{bf}
}}}\\
\\

Both aspects of complementarity are a consequence of the modes of the photons. Thus it is the wave nature of the photons and not the particle-like aspect which is responsible for complementarity. The indivisible energy packets of the single photons are only the reason for occupied (excited) or not occupied (not excited) modes.\\

To investigate the physical background of these aspects of the complementarity principle a set of new experiments was performed to show in detail the properties of single photons in certain modes.\\

\section{Spontaneous parametric down conversion (SPDC) as a tool}

\begin{figure}[h]
\centering
\includegraphics[width=9cm]{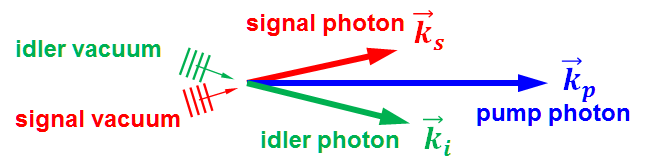}
\caption{Scheme of the emitted light in SPDC: The entangled signal (red) and idler (green) photons occur at opposite sides of the cone.}
\label{Fig1}
\end{figure}

In spontaneous parametric down conversion an optically transparent crystal with a second order nonlinearity ($\chi^{(2)}$) is used to generate two entangled photons from one pump photon. In this process energy and momentum is conserved. By choosing a suitable orientation of the crystal phase matching between the wave of the pump photon and the two waves of the newly produced photons can be realized and thus the efficiency of the process enlarged significantly \cite{ref:Boyd}, \cite{ref:Photonics} and \cite{ref:Walborn}. The two new photons are usually named signal and idler photons. In type I phase matching which is realized by a special orientation of the crystal the signal and the idler photons belong to the same light cone structure which is emitted cylindrically symmetric around the pump photon direction as illustrated in figure \ref{Fig1}.

Because of momentum conservation the two photons appear on opposite sides of the cone. As a result of the entanglement of the two photons including their temporal synchronization one of the photons (e.g. the idler photon) can be used as a reference for the other photon (then the signal photon). This allows in addition to usual single photon measurements coincidence measurements from the simultaneous observation of two clicks. Because of the entanglement (and correlation) properties of the signal photon can be determined from measuring the correlated properties of the idler photon. As will be shown below this allows new measurements regarding the investigation of the complementarity principle.

But even more important information is available from using spontaneous parametric down conversion for the investigation of complementarity. In the quantum description of the generation of the new photons the 3-wave mixing of the pump mode with the vacuum modes is applied.

The electric field components of the signal and the idler field are given by equation 1 and 2.

\bea
E_{signal}^{(+)}(\bold r, t)&=&a_{s0}(\bold r, t)+Ca_{pump}(\bold r, t)\ad_{i0}(\bold r, t)
\label{th1}
\eea
\bea
E_{idler}^{(+)}(\bold r, t)&=&a_{i0}(\bold r, t)+Ca_{pump}(\bold r, t)\ad_{s0}(\bold r, t)
\label{th2}
\eea

\begin{figure}[h]
\centering
\includegraphics[width=9cm]{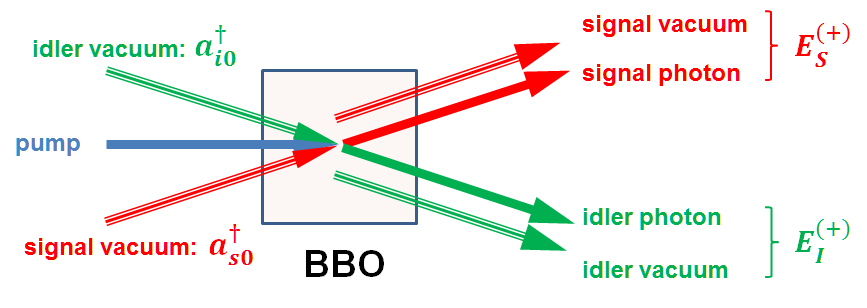}
\caption{Scheme of 3 wave mixing process with the vacuum field contribution}
\label{Fig2}
\end{figure}

With this simplified model as used in \cite{ref:Milonni} and \cite{ref:H_3crystal} an effective Hamiltonian is applied. The couplings are described by $a_{pump}\ad_{i0}$ and $a_{pump}\ad_{s0}$, where $a_{pump},a_s$ and $a_i$ ($\ad_{pump},\ad_s$ and $\ad_i$) are photon annihilation (creation) operators for the pump, signal, and idler fields. This way the annihilation (creation) of the pump photon and the simultaneous creation (annihilation) of signal and idler photons in the SPDC 3-wave mixing process with the vacuum field contributions ${a_{s0}}$ as signal vacuum and ${a_{i0}}$ as idler vacuum is considered. The electric fields leaving the crystal are the sum of the created photon fields and the undisturbed vacuum fields crossing the crystal. This description of the SPDC process is illustrated in figure \ref{Fig2}.

The important observation here is that in this spontaneous parametric down conversion the vacuum fields are directly involved in the generation process of the two new photons. This allowed the investigation of the physical background of complementarity with new experiments using SPDC as a tool. By measuring a single signal photon or single idler photon these vacuum fields will not play any role. But if the two photons are measured in coincidence these vacuum fields will mix in the formulas of the measured intensity as Glauber described in his photodetection theory of the measuring process \cite{ref:Glauber} and thus entanglement is provided.

In summary, spontaneous parametric down conversion (SPDC) allows a detailed investigation of the influence of the vacuum fields in the generation process of single photons and in the entanglement process. This will be, as shown below, a very powerful tool to investigate complementarity.

\section{Induced coherence in the 3-crystal set up}

\begin{figure}[h]
\centering
\includegraphics[width=10cm]{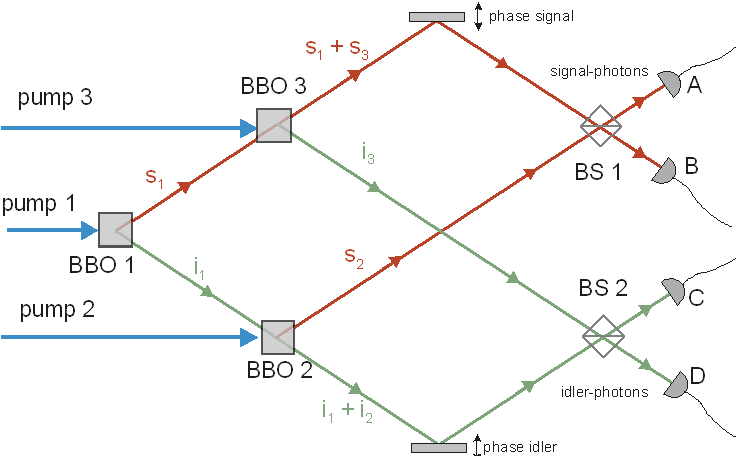}
\caption{Scheme of the induced coherence setup with 3 crystals as published and described in detail in \cite{ref:H_3crystal}.}
\label{Fig3}
\end{figure}

As published in \cite{ref:H_3crystal} the addition of the third crystal in the known induced coherence setups  \cite{ref:Mandel_ind} and with higher visibility described in \cite{ref:H_ind} allows a very direct observation of the randomness of the vacuum fields. The experimental setup is shown in figure \ref{Fig3}.

\begin{figure}[h]
\centering
\includegraphics[width=14cm]{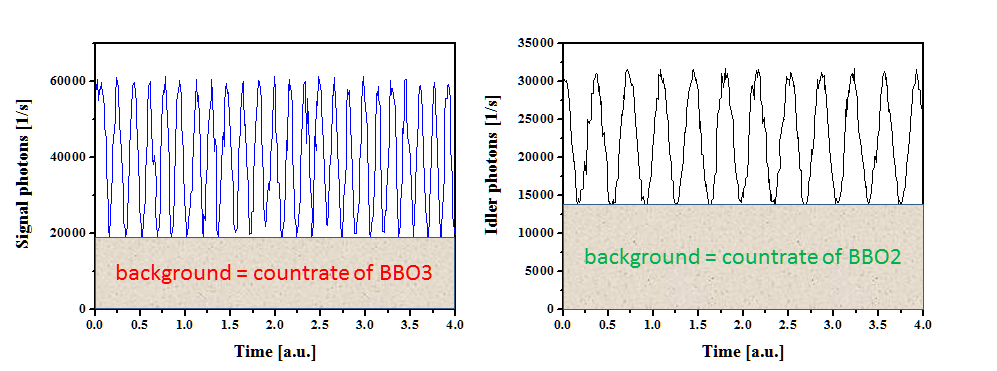}
\caption{Experimental results of the interference fringes in the 3-crystal induced coherence measurement of figure \ref{Fig3} as published and described in \cite{ref:H_3crystal} in detail. On left the count rate of the signal photons measured with detector A and on right  the count rate of the idler photons measured with detector D is shown. Although the signal photons of crystal BBO1 and BBO3 are perfectly coherent all photons of crystal BBO2 are not coherent and produce the background signal in this figure on left. The same argument is valid for the idler photons at detector D.}
\label{Fig4}
\end{figure}

In this set up the crystals BBO 1 and BBO 2 are aligned for the same idler mode and crystal BBO 1 and BBO 3 are aligned for the same signal mode. Therefore the two signal photons s1 and s2 are perfectly coherent as well as the idler photons i1 and i3 if the whole setup is aligned well. As a result at detector A perfect fringes with visibilities of almost 1 could be measured if crystal BBO 3 was not pumped and in the same way with detector D visibilities of almost 1 could be detected if crystal BBO 2 was not pumped. But if all three crystals are pumped, simultaneously, both interference diagrams show an incoherent background as it is shown in figure \ref{Fig4}.

\begin{figure}[h]
\centering
\includegraphics[width=12cm]{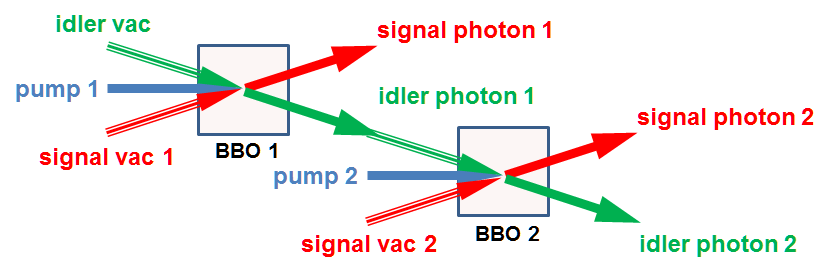}
\caption{Scheme of the induced coherence setup with the vacuum field contribution.}
\label{Fig5}
\end{figure}

As it was shown in reference \cite{ref:H_3crystal} this result can be explained using the set of equations as given in equations 1 and 2 from above. The result is illustrated in figure \ref{Fig5}.\\

As can be seen from this figure and also shown in the equation 3 the fields of the signal photon 1 and of the signal photon 2 are coherent because in the generation process of these two photons the pump is a coherent state and therefore the fields of pump 1 and pump 2 have a fixed phase relation. The idler vacuum responsible for the 3-wave mixing in the generation process of the two photons is the same. Therefore the 3-wave mixing process in the two crystals BBO1 and BBO2 is realized with the same phase relations and thus the two signal photons are generated with the same phase as can be seen in equation 3 (of course the path delays between the two crystals has to be recognized).

\bea
E_{S-BS}^{(+)}(\bold r, t)&=&a_{s10} +ia_{s20}e^{i \phi_{s2}}+C_1a_{pump1} \ad_{i0} +iC_2 a_{pump2} \ad_{i0} e^{i \phi_{s2}}
\label{th1}
\eea

The situation for the generation of the two idler photons, idler photon i1 and idler photon i2, is completely different as illustrated in figure \ref{Fig5} and in equation 4.

\bea
E_{I-BS}^{(+)}(\bold r, t)&=&a_{i10} +ia_{i20}e^{i \phi_{i2}}+C_1a_{pump1} \ad_{s10} +iC_2 a_{pump2} \ad_{s20} e^{i \phi_{i2}}
\label{th1}
\eea

Although the two idler photons from the two crystals BBO 1 and BBO 2 are excitations of the same spatial vacuum mode as already shown in figure \ref{Fig5} the generation process of the two idler photons is a 3-wave mixing of the pump which is again coherent but of two clearly distinguishable signal vacuum fields signal vac1 and signal vac2. These two vacuum fields are of course not at all coherent (which is also demonstrated by the result of the experiment, see figure \ref{Fig4}, right). As result the idler photon i1 and the idler photon i2 have a random phase relation. Thus in the induced coherence experiment with three crystals as shown in figure \ref{Fig3} for the idler photons detected with detector D a coherent signal can be observed for the photons produced in crystal BBO 1 or BBO 3. But all the idler photons generated in crystal BBO 2 are not coherent neither to the photons from BBO 1 nor  to the photons from BBO 3. As result all the idler photons generated in BBO 2 appear as an incoherent background in the interference measurement at detector D as observed in figure \ref{Fig4}.

The same argumentation is valid for the measurement at detector A. All the signal photons generated in crystal BBO 3 are not coherent to the signal photons generated in crystal BBO 1 or in crystal BBO 2. As result all the signal photons from crystal BBO 2 appear as incoherent background in the interference measurement at detector A.

Regarding the question of complementarity from these experimental results it can be summarized:

\begin{bf}
Photons generated by the same vacuum field will be coherent if all other participating fields are also coherent. But if the photons are generated by different vacuum fields they are not coherent.
\end{bf}

This is also true even if the two photons belong to the same spatial TEM$_{00}$ mode as it was applied in the experiments above for both the idler channel i1 and i2 as well as the signal channel s1 and s3. This means it is possible to generate single photons in the same spatial mode but with different phases. These photons are, as in the experiment above, also distinguishable. In the described measurement they could easily be distinguished by measuring the related signal photon s1 and s2 as reference for the idler photons i1 and i2 or the idler photons i1 and i3 for the signal photons s1 and s3 in coincidence.

In summary of these experimental results it can be concluded:\\

{\centering
\fbox{\parbox{\linewidth}{
\centering
\begin{bf}
The randomness of the vacuum fields causes temporal complementarity.\\
\end{bf}
}}}

\section{Stimulated coherence}

\begin{figure}[h]
\centering
\includegraphics[width=11cm]{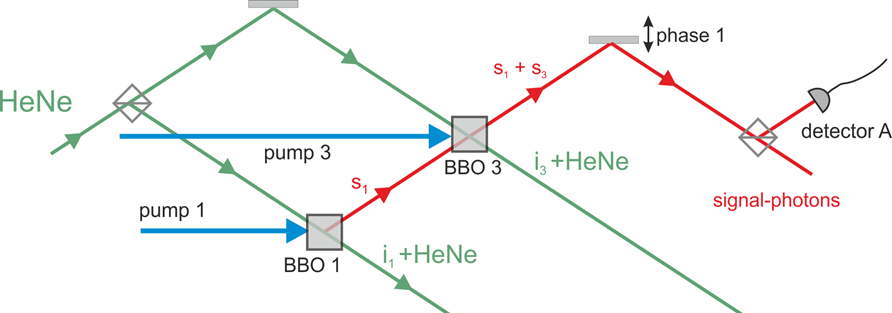}
\caption{Scheme of of the setup with two sequential BBO-crystals for measuring stimulated coherence overwriting the random vacuum field as described in detail and published in \cite{ref:H_stim}.}
\label{Fig6}
\end{figure}

The random phase of a vacuum field mode can be overwritten by the coherent mode of a laser \cite{ref:Mandel_stim} as described in detail in \cite{ref:H_stim}. In our experiment the radiation of a HeNe-laser was used as stimulating light field in the 3-wave mixing process inside the SPDC-crystal in the same TEM$_{00}$ mode as it was used in the idler channels of the two separated crystals BBO1 and BBO3 in the induced coherence experiment above (see figure \ref{Fig3}). The experimental scheme is depicted in figure \ref{Fig6}.
 
\begin{figure}[h]
\centering
\includegraphics[width=7cm]{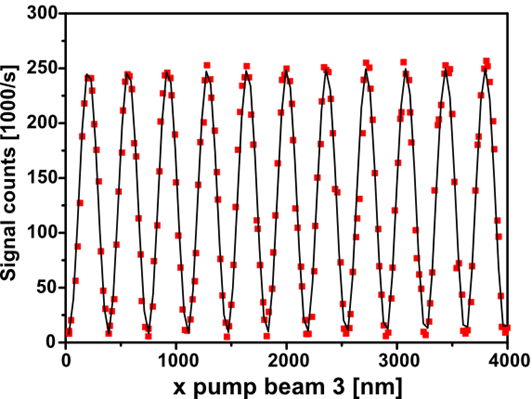}
\caption{Interference of single signal photons emitted from crystals BBO1 or BBO3 of figure \ref{Fig6} while changing the delay between pump 1 and pump 3 resulting in a visibility of 95\% as published in \cite{ref:H_stim}}.
\label{Fig7}
\end{figure}

As can be seen from the scheme of figure \ref{Fig6} the two crystals BBO1 and BBO3, which are positioned in the same way as in figure \ref{Fig3}, are coupled via the signal channel s1 and s3. As in the experiment of figure \ref{Fig3} these two signal channels would be not coherent without the HeNe-laser radiation as a consequence of the random phase of the two distinguishable vacuum fields in the idler channels i1 and i3. But by overwriting these two idler vacuum channels with the coherent light field of the HeNe-laser the phases of the two signal channels are fixed and an interference with a high visibility of 0.95 was observed at detector A if one of the pump laser radiation is delayed as shown in figure \ref{Fig7}  \cite{ref:H_stim}.
 
In this case the remaining spontaneous emission triggered by the vacuum fields appears as an incoherent background in this measurement. But their count rate is so low that it is almost not visible in the single photon interference pattern.

\begin{figure}[h]
\centering
\includegraphics[width=11cm]{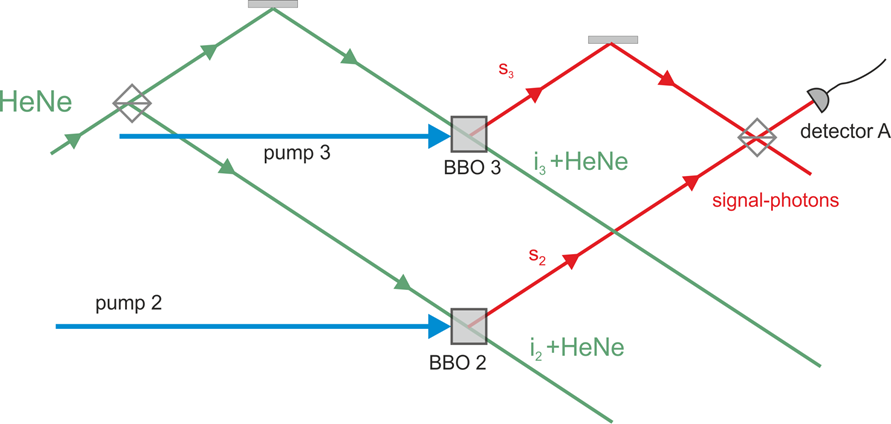}
\caption{Scheme of the setup with two parallel pumped BBO-crystals for measuring stimulated coherence by overwriting the random vacuum field as described in detail and published in \cite{ref:H_stim}}
\label{Fig8}.
\end{figure}

This effect can also be demonstrated by using two crystals in a parallel arrangement as it is also described in detail in \cite{ref:H_stim}. The two SPDC-crystals BBO2 and BBO3 are pumped coherently from the same pump laser source as in all other experiments before and the HeNe-laser radiation is applied in the idler channels of the two crystals. The experimental scheme is depicted in figure \ref{Fig8}.

\begin{figure}[h]
\centering
\includegraphics[width=7cm]{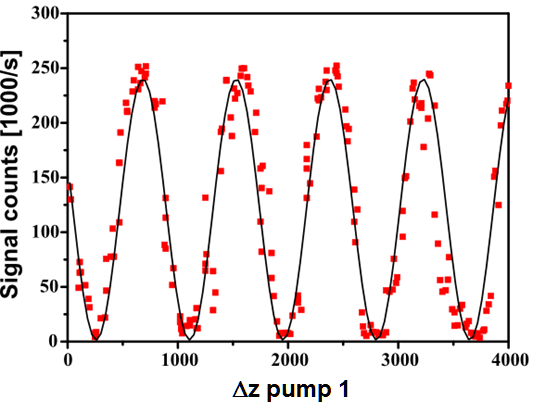}
\caption{Interference measurement between the single signal photons from crystals BBO2 or BBO3 of figure \ref{Fig8} while changing the delay between pump 2 and pump 3 resulting in a visibility of 98\% as published in \cite{ref:H_stim}}
\label{Fig9}
\end{figure}

If a delay line is inserted in the beam of pump 3 it is possible to observe single photon interference between the two signal channels s2 and s3 at detector A. The result of this measurement is given in figure \ref{Fig9}.\\

As can be seen from this figure \ref{Fig9} the single photon count rate for the signal photons provided from crystals BBO2 or BBO3 shows clear fringes as a function of the delay of the pump beam with high visibility of 98\%. Without the HeNe-laser radiation no single photon interference could be observed at all. Again the remaining spontaneous emission triggered by the random vacuum fields appears as an incoherent background in this measurement, but with very low count rate that it is almost not visible in this measurement.\\

Summarizing this part of the discussion it can be stated:\\

{\centering
\fbox{\parbox{\linewidth}{
\centering
\begin{bf}
The random vacuum fields can be “overwritten” by coherent light fields, e.g. from laser sources.\\
\end{bf}
}}}
\\

\section{Complementarity in the spatial dimension}

\begin{figure}[h]
\centering
\includegraphics[width=9cm]{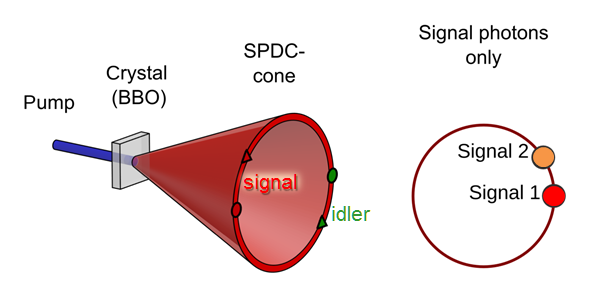}
\caption{Scheme of the spatial emission of spontaneous parametric down conversion (type I) of a BBO crystal and the position of the two TEM$_{00}$ signal photon modes signal 1 and signal 2 which are selected by the detection system and superimposed for investigating the lateral coherence between them \cite{ref:CoVol}.}
\label{Fig10}
\end{figure}

After discussing the complementarity for single photons in the temporal dimension the next step is their investigation in the spatial dimension as described in \cite{ref:CoVol}. While above the temporal coherence of single photons and their distinguishability were analyzed now the coherence of single photons which are potentially belonging to two or more spatial modes will  be discussed. Therefore single photon interference was measured for photons in transversally distributed modes. These modes are superimposed and the resulting visibility and distinguishability of these modes is measured.\\
For this purpose an experiment was set up in which spontaneous parametric down conversion (type I) was used as a tool again. The scheme is depicted in figure \ref{Fig10}.

\begin{figure}[h]
\centering
\includegraphics[width=13cm]{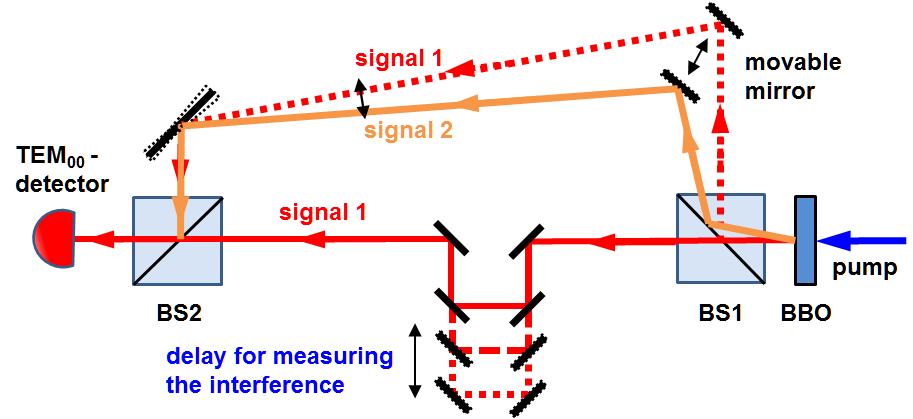}
\caption{Scheme of the Mach-Zehnder-interferometer for measuring the interference of different positions of the selected TEM$_{00}$ modes of signal 1 and signal 2 within the cone.}
\label{Fig11}
\end{figure}

In this case a single light cone is emitted behind the crystal with a cylindrical symmetry around the pump beam direction. The single idler and signal photons appear on opposite sides of the cone. Our spatially selective TEM$_{00}$ mode detector was set up to measure a small emission share from this cone in the areas of signal 1 and signal 2 as depicted in figure \ref{Fig10}. The resulting two TEM$_{00}$ modes of signal 1 and signal 2 were fed into a Mach-Zehnder interferometer as shown in figure \ref{Fig11}.\\

\begin{figure}[h]
\centering
\includegraphics[width=8cm]{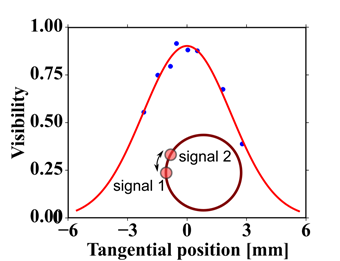}
\caption{Measured visibility of the interference fringes of the two TEM$_{00}$ modes of signal 1 and signal 2 as a function of their tangential distance along the emitted cone structure.}
\label{Fig12}
\end{figure}

With this arrangement different TEM$_{00}$ modes of the emitted single photons could be overlaid using a delay line for zero delay and the interference visibility could be measured as a function of the distance of the two TEM$_{00}$ modes within the cone as shown in figure \ref{Fig10} right. The result of this measurement is shown in figure \ref{Fig12}.

The maximum visibility in this case was determined to about 90\% as a consequence of no corrections for any background or other disturbing signals. The experimentally observed width of the visibility curve demonstrated that only photons inside the TEM$_{00}$ detection mode are coherent as described in detail in \cite{ref:CoVol}.

\begin{figure}[h]
\centering
\includegraphics[width=8cm]{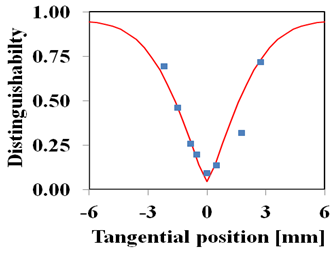}
\caption{Distinguishability between the single photon in the TEM$_{00}$-modes signal 1 or signal 2 as a function of the tangential distance of signal 2 in relation to the signal 1 (\cite{ref:CoVol}).}
\label{Fig13}
\end{figure}

The also available idler photon on the opposite side of the light cone (see figure \ref{Fig10}, left) can be used as a reference for the signal photon. From the coincidence measurement of the idler and the signal photons the distinguishability D between the two TEM$_{00}$ measured modes could be determined. The result of this measurement is given in figure \ref{Fig13}.

From these two measurements at each distance between the two TEM$_{00}$ modes signal 1 or signal 2 the complementarity value of $D^2+ V^2$ was determined to values between 0.8 and 0.9 \cite{ref:CoVol}. These values confirm that in this measurement the complementarity principle in the spatial dimension is nicely fulfilled. The reason why the sum is slightly below 100\% is a consequence of the not corrected visibility measurement of figure \ref{Fig12}.\\

These experimental results can be summarized in the following way:\\

\begin{bf}
- Photons in a TEM$_{00}$ mode are only coherent from the same mode.

- These photons are not distinguishable.\\
\end{bf}

Regarding complementarity in the spatial dimension it can be concluded:\\

{\centering
\fbox{\parbox{\linewidth}{
\centering
\begin{bf}
Distinguishable spatial modes are generated from different vacuum fields. Because of the randomness of the vacuum fields these modes are not coherent. Thus also in the spatial dimension complementarity is a consequence of the randomness of the vacuum fields.\\
\end{bf}
}}}
\\

\section{Complementarity for single photons in a higher order spatial modes}

\begin{figure}[h]
\centering
\includegraphics[width=11cm]{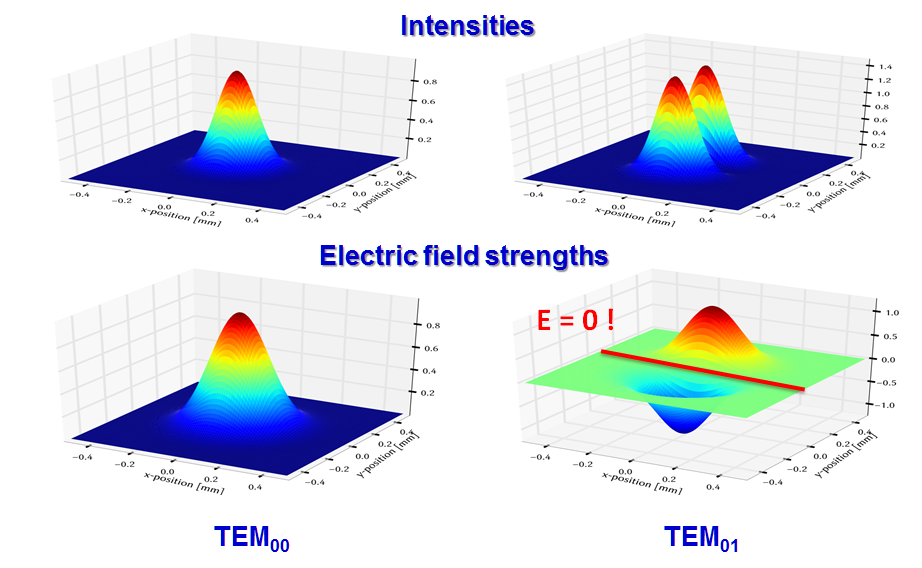}
\caption{Calculated intensity (top) and electric field patterns (bottom) for a TEM$_{00}$ (left) and a TEM$_{01}$ (right) mode. Both the electric field and the intensity of the TEM$_{01}$-mode is zero between the two humps.}
\label{Fig14}
\end{figure}

After having the mode concept applied for single photons in the fundamental mode, successfully, the question about complementarity for photons in higher order spatial modes will be asked. These modes show intensity structures with at least two or more humps as e.g. the TEM$_{01}$ Gauss-Laguerre or Gauss-Hermite modes. The electric field has opposite phases in the neighboring humps and between the humps both the electric field and the intensity is 0 (see figure \ref{Fig14}). 

This suggests the questions: Does a single photon in such higher order mode show interference as the classical mode would do and can the single photon be localized in one of the humps?

For investigating the complementarity principle for single photons in such higher order spatial modes reference photons are needed to determine the distinguishability as in all previous measurements described above. The known way to realize such a photon correlation is spontaneous parametric down conversion as also described above.
 
\begin{figure}[h]
\centering
\includegraphics[width=12cm]{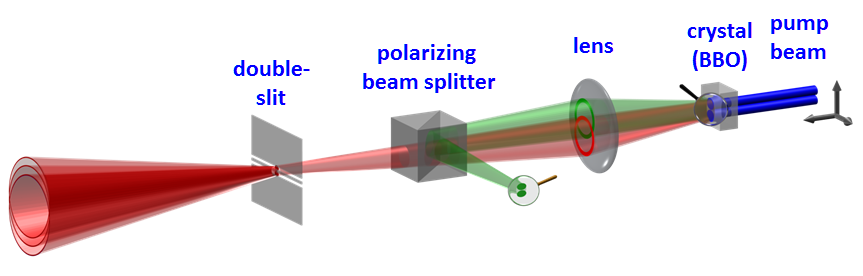}
\caption{Scheme of the double slit experiment using a TEM$_{01}$ mode to pump at type II SPDC crystal as described in detail and published in \cite{ref:DS1}. The two humps of the signal photons were imaged onto the double slit and the idler photons coupled out with a polarizing beam splitter and imaged on the idler detector for reference.}
\label{Fig15}
\end{figure}

Therefore we applied photons in a TEM$_{01}$ mode to pump at type II SPDC crystal \cite{ref:DS1}. Indeed the out coming signal and idler photons showed a double hump structure in the near field of the crystal similar as the pump photon. Therefore the single signal photons in this TEM$_{01}$-like mode near field structure were used in the Young’s double slit experiment as shown figure \ref{Fig15} and described in detail in \cite{ref:DS1}.

The two humps of the signal photon light were imaged onto the two slits. With the idler photon detector as reference "which-slit" information based on the near field correlation could be measured in coincidence if the signal photon detector was placed right behind the double slit. A very strong correlation of more than 95\% was found for the two photons emitted both from the same hump of the pump mode. We observed this result also for single photon pairs in higher order modes as e.g. TEM$_{02}$ or TEM$_{11}$ modes.

While imaging the single signal SPDC-photons onto the double slit as shown in figure \ref{Fig15} and described in detail in \cite{ref:DS1} interference fringes with good visibility in the range of 0.6 could be observed in the far field behind the slits. This result seem to conflict with the complementarity principle. But this simple arrangement of figure \ref{Fig15} includes an unfair sampling for the measured photons. Much more photons are detected for the "which-slit" information compared to the photons observed in the interference measurement as it was discussed in reference \cite{ref:Leach}. 

A detailed analysis showed that this simple experiment shows a reach variety of transversal light structures behind the SPDC crystal as a consequence of the TEM$_{01}$ pump mode \cite{ref:DS2} and the resulting complex  phase matching conditions. Unfortunately the higher order pump mode does not directly transform to the same down converted mode of the signal and idler photons.

But based on the results of this complicated and full theoretical description of the SPDC process without any fitting parameters \cite{ref:DS2} and the realized almost perfect match of the experimental results it was possible to modify the experimental setup with apertures in the far field between the crystal and the polarizing beam splitter (near the lens in figure \ref{Fig15}) to avoid the unfair sampling. Both the idler and the signal photon detector are measuring the same photon pairs in the near and in the far field of the signal photons, then. Thus which-slit information and interference visibility can directly be combined in this modified setup.

\begin{figure}[h]
\centering
\includegraphics[width=12cm]{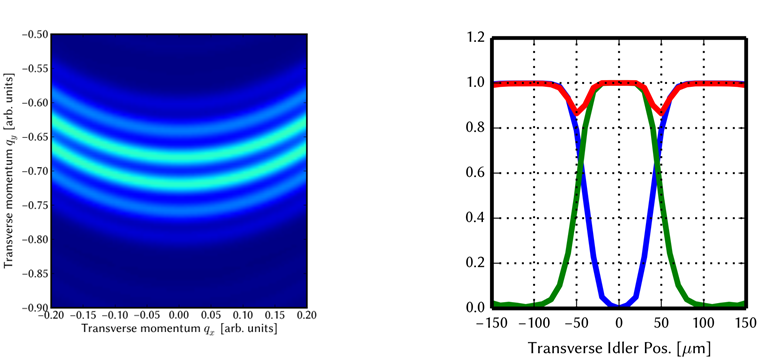}
\caption{Result of the double slit interference with signal photons in a TEM$_{01}$-like mode (left) and the associated distinguishability and visibility measurements (right) as described in detail in \cite{ref:DS3} with the setup of figure \ref{Fig15} but with apertures for fair sampling.}
\label{Fig16}
\end{figure}

This way it was possible to obtain the double slit interference with signal photons in a TEM$_{01}$-like mode to our knowledge for the first time. The result as described in detail in \cite{ref:DS3} is shown in figure \ref{Fig16}.

As can be seen from the left side of figure \ref{Fig16} the far field interference pattern of the signal photons shows on this side of the emitted light cone a dip in the middle of the interference structure. This indicates that the electric field of the two humps imaged onto the two slits has a phase shift of 180° because otherwise the maximum of the interference would be obtained in the middle of that structure. So far the TEM$_{01}$-like mode structure is confirmed from this observation.\\

On the right side of figure  \ref{Fig16} the distinguishability D, the visibility V and the $D^2+ V^2$ values of the signal photons are given as a function of the vertical position of the idler detector in this coincidence measurement. If the idler detector is positioned in the middle of the TEM$_{01}$ mode the visibility is as high as possible (V = 1) and the distinguishability D of the signal photon passing slit 1 or slit 2 is zero. If the idler detector is moved outside of the middle of the mode the visibility drops and the distinguishability increases. The $D^2+ V^2$ value is 1 in maximum which is expected and has little drops in between.\\

Therefore also for single photons in a higher order transversal mode (in our case  TEM$_{01}$) the complementarity principle is fulfilled. But in the context of the previous discussions the physical background of this behavior can be analysed, too. First it can be concluded that even single photons in a higher order transversal mode show the classically expected interference probability structure. Although the near field coincidence measurement of the signal and idler photon pair shows a strong localization of the emission spot of the bi-photon in just one of the humps the interference pattern is given by the full mode.\\

It can be concluded from this experiment that also the higher order transversal modes exist as non-occupied vacuum field modes. With this physical background there is a very intuitive explanation of the result of figure \ref{Fig16}:

As long as our detectors see the vacuum field contributions in the generation process of the bi-photons only in the desired TEM$_{01}$ mode the obtained visibility V is 1 as a consequence of the coherence of this mode. But as soon as the idler detector is moved towards one of the humps photons not belonging to the desired TEM$_{01}$ mode are measured. These photons are generated by different vacuum fields. These vacuum fields are not coherent to the vacuum fields of the first mode. Therefore the visibility drops. But then the two modes become distinguishable and the distinguishability D in this measurement increases. In any case the $D^2+V^2$ value is maximum one.\\

Finally these measurements can be summarized in the following way:\\

{\centering
\fbox{\parbox{\linewidth}{
\centering
\begin{bf}
Photons in a higher order transversal modes are coherent and not distinguishable and vice versa ($D^2+V^2$ = 1) .
\\
\end{bf}
}}}
\\
\\
\section{Conclusion}
In summary of the here discussed experiments the complementarity principle was demonstrated as a consequence of the measuring process, which makes a selection of the mode function and the involved vacuum field contributions. Finally this may give a new conceptual perspective towards quantum optics.

The physics behind the complementarity principle may be summarized as follows:\\

\begin{bf}
- uncertainty is a consequence of the wave nature of quantum objects

- duality is a consequence of the energy packet in this wave (mode)

- the detection system selects coherent or distinguishable modes

- vacuum field modes are random: intrinsic randomness in QM

- coherent (laser) modes can overwrite the random vacuum fields
\end{bf}\\

The most important consequence of this discussion is:\\

{\centering
\fbox{\parbox{\linewidth}{
\centering
\begin{bf}
Only modes based on the same vacuum are coherent.
\\
\end{bf}
}}}
\\

Therefore each experimental situation can be analyzed for the measured modes and their vacuum contributions. Complementarity means that photon modes based on the same vacuum will be coherent because the excitation of these modes does not change their phase. But as in the example of the 3-crystal experiment the same spatial mode can be excited by different vacuum field contributions and then the single photon modes can be incoherent.  Therefore the final consequence of this discussion is:\\

{\centering
\fbox{\parbox{\linewidth}{
\centering
\begin{bf}
The physics of the vacuum field modes causes complementarity in quantum optics.
\\
\end{bf}
}}}
\\

\acknowledgments
Detailed and long standing discourses as well as fruitful scientific collaborations over the years with Axel Heuer, Peter Milonni, Martin Ostermeyer, Wolfgang Schleich and Martin Wilkens and with the members of my group are greatfully acknowledged. In addition very helpful have been the discussions with Bob Boyd, Claude Fabre, Berthold Englert, Gerd Leuchs, Pierre Meystre, Harry Paul, Marlan Scully and Ewan Wright and many others.

\end{document}